\PassOptionsToPackage{table,usenames,dvipsnames}{xcolor}

\documentclass[%
aps,
reprint,
 amsmath,amssymb,
prper,
floatfix
]{revtex4-2}

\usepackage{multirow}
\usepackage{graphicx}% Include figure files
\usepackage{dcolumn}% Align table columns on decimal point
\usepackage{bm}% bold math
\usepackage{tikz}
\usepackage{pgfplots}
\usepackage[export]{adjustbox}
\pgfplotsset{compat=1.17}
\usepackage{listings}
\begin{document}

%\preprint{APS/123-QED}

\title{When AI Evaluates Its Own Work:\\ Validating Learner-Initiated, AI-Generated Physics Practice Problems}

\author{Tobias Geisler}
 \email{dblthere@gmail.com}
 \affiliation{%
Department of Information Technology and Electrical Engineering, ETH Zurich, 8092 Zurich, Switzerland
}%

\author{Gerd Kortemeyer}
 \email{kgerd@ethz.ch}
 \affiliation{%
Rectorate and ETH AI Center, ETH Zurich, 8092 Zurich, Switzerland
}%
\altaffiliation[also at ]{Michigan State University, East Lansing, MI 48823, USA}

\date{\today}

\begin{abstract}
Large language models (LLMs) can now generate physics practice problems in real time, yet the educational value of these items hinges on rapid, reliable post-generation vetting.  In this exploratory study, we investigated which automated checks are both technically feasible and pedagogically meaningful when exercises are produced on demand within a chatbot interface.  A cohort of 34 introductory-physics students generated and attempted 543 practice problems during exam preparation. Each item was labeled by an expert on a wide range of quality attributes and presented to the learners in pairs to record their preference.  We then (i) benchmarked three commodity LLMs as ``judges'' against the expert labels, (ii) quantified which attributes predict student choice via random-forest models, and (iii) triangulated these results with free-form exit surveys. Only a small subset of the original metric items proved necessary to reliably address student preferences either directly or by proxy. The study demonstrates that scalable formative assessment does not require exhaustive scoring: a carefully curated core of structural and learner-visible checks is sufficient to ensure both technical soundness and user appeal.  The findings provide a practical blueprint for deploying real-time, AI-generated practice in physics and other quantitative disciplines.

\end{abstract}

\maketitle

\section{Introduction}
\subsection{Motivation}
Physics instructors routinely observe that students ask for additional practice problems on particular topics, especially in the run-up to exams. Students may have worked through end-of-chapter exercises, old exams, and other available formative assessments, yet still struggle with specific concepts and feel unprepared for upcoming summative assessments. In large-enrollment courses, instructors typically cannot respond to individualized, topic-specific requests --- and they certainly cannot write new practice problems on demand in a scalable way. 

\begin{figure}
\begin{center}
\includegraphics[width=\columnwidth]{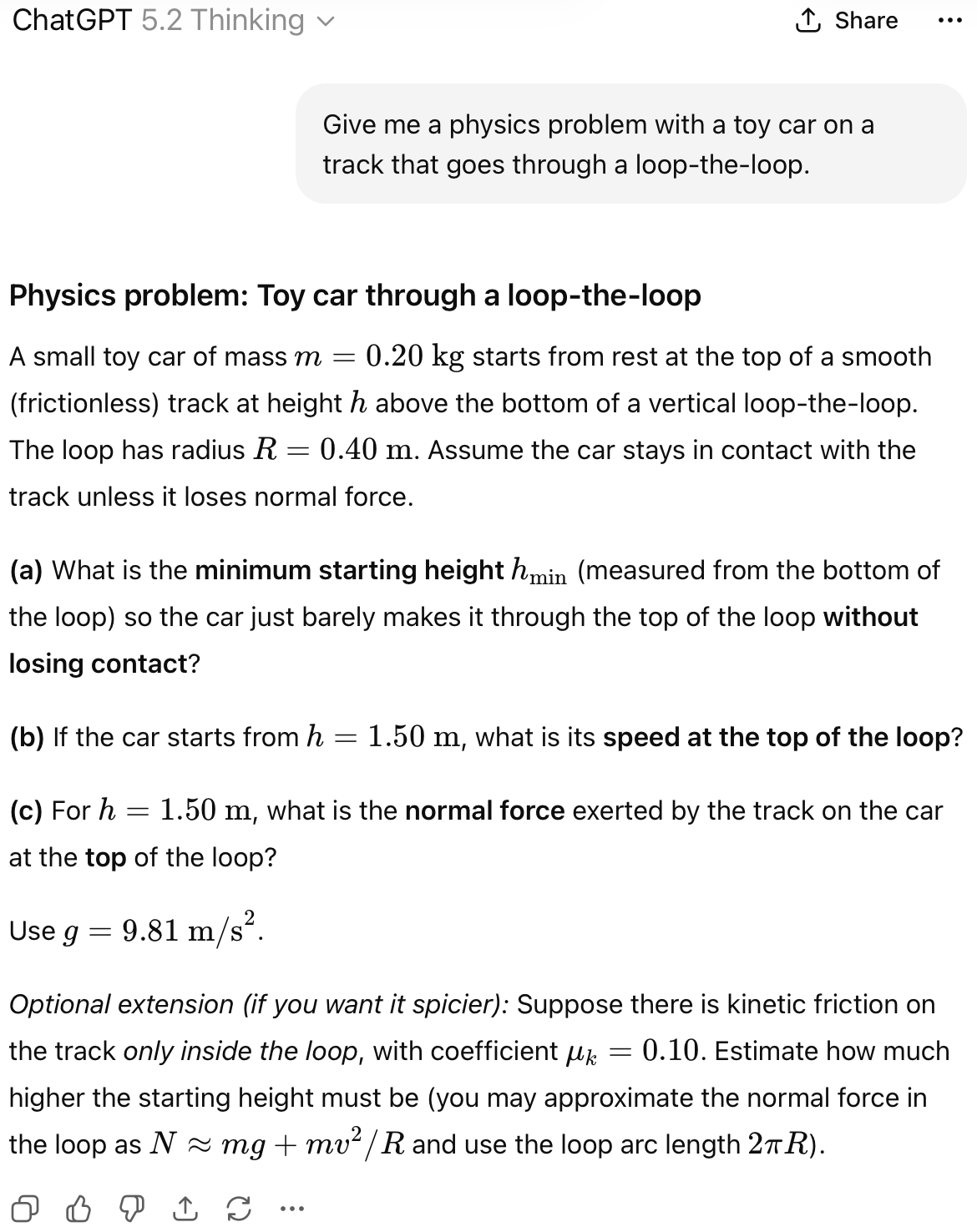}
\end{center}
\caption{An example of an on-demand problem generated by GPT-5.2 Thinking.}
\label{fig:example}
\end{figure}

In such situations, students who can afford it may turn to private tutoring. In principle, generative AI offers a more equitable alternative: students can ask a chatbot to generate practice problems on demand. In practice, however, when these systems lack access to course materials, they may produce problems that are misaligned with the course, underspecified or internally inconsistent, or accompanied by incorrect solutions. Figure~\ref{fig:example} illustrates this failure mode using GPT-5.2 Thinking~\cite{gpt52}. The dialogue begins with a standard, solvable problem: the scenario and questions are clearly formulated, all required information is present (and the value of $g$ is only specified in part (c), where it is first needed), and the chosen height $h$ indeed keeps the car on track. When asked for solutions, the system also computes parts (a)--(c) correctly. We would therefore consider parts (a)--(c) ``valid'' in a physics sense. However, the final optional part contains a serious physical error: the claimed ``approximation'' $N \approx mg + mv^2/R$ is not justifiable in this context. When asked to justify it, the system argued that the approximation is needed to make the integral easier --- which is true, but we would argue that the problem is physically ``invalid'' with that misleading additional information.

In short, for a variety of reasons, AI-generated problems may be unsuitable for practice even before pedagogical considerations enter the picture. An unsolvable, contradictory, or physically incorrect problem has little formative value (unless diagnosing its flaws is itself an intended learning objective~\cite{shekoyan2009using,kagan2019problem}). In this study, we therefore investigate metrics that can be used to verify the ``nuts-and-bolts'' validity of on-demand generated physics practice problems. Using problems generated by student prompt during a simulated study session as samples, we evaluate candidate validation metrics by their agreement with expert judgment and by their relationship to surface-level student expectations.

\subsection{Background}
Problem solving is an essential component of physics teaching and learning~\cite{reif1976,arons1996teaching,national2000}. Besides summative assessments like quizzes and exams, formative assessments such as homework and exercise sheets also play a central role in learning physics~\cite{dufresne2004,wagner2012promoting}.  To provide rapid feedback,
interactive online homework has proven effective in learning physics across a number of studies~\cite{kashy03,kortemeyer08,risley2001,tang2002,bonham2003,gutmann2018}. These materials, however, are instructor-provided and may not address the individual needs and desires of the learners (an alternative are student-authored problems, which may vary in quality~\cite{bates2014assessing}).

Artificial intelligence (AI) is increasingly being embraced in physics education~\cite{sperling2024artificial,wattanakasiwich2025physics,kilde2025generative}, offering promising applications for both teaching and learning physics. Due to growing corpora of training materials and improved reasoning algorithms, Large language models (LLMs) have developed strong conceptual reasoning capabilities~\cite{kortemeyer2025multilingual} (even though their conceptual framework might not always be stable~\cite{dunlap2025descending}). They have thus, proven valuable for instructors, helping in the creation of tailored materials and tasks~\cite{kuchemann23,lademann2025augmenting,wattanakasiwich2025physics}, providing personalized feedback to students~\cite{bitzenbauer2023,kortemeyer2024ethel},  assisting in grading~\cite{kortemeyer2025assessing,chen2025grading}, and training of teaching staff~\cite{gregorcic24}. Students report that these models represent always-available resources, which appear to be non-judgmental, imperturbable, and ever-patient~\cite{vasconcelos2023,ding2023students,balabdaoui2024survey}.  

While AI-models have shown remarkable performance in solving physics problems in a number of studies~\cite{kortemeyer23ai,pimbblet2024,polverini24,kortemeyer2025multilingual}, generating such assessment materials is an active area of research. Despite their potential, the quality of these automatically generated problems varies widely~\cite{niu2023exercise,Maity2024How}, sometimes leading to student confusion and misconceptions~\cite{kirschner2006why}. Thus, in most current studies, instructors use the AI system to generate the problems, and then the instructors themselves judge which generated problems they present to the students~\cite{el2024streamlining}.

\begin{figure*}
\begin{center}
\includegraphics[width=0.8\textwidth]{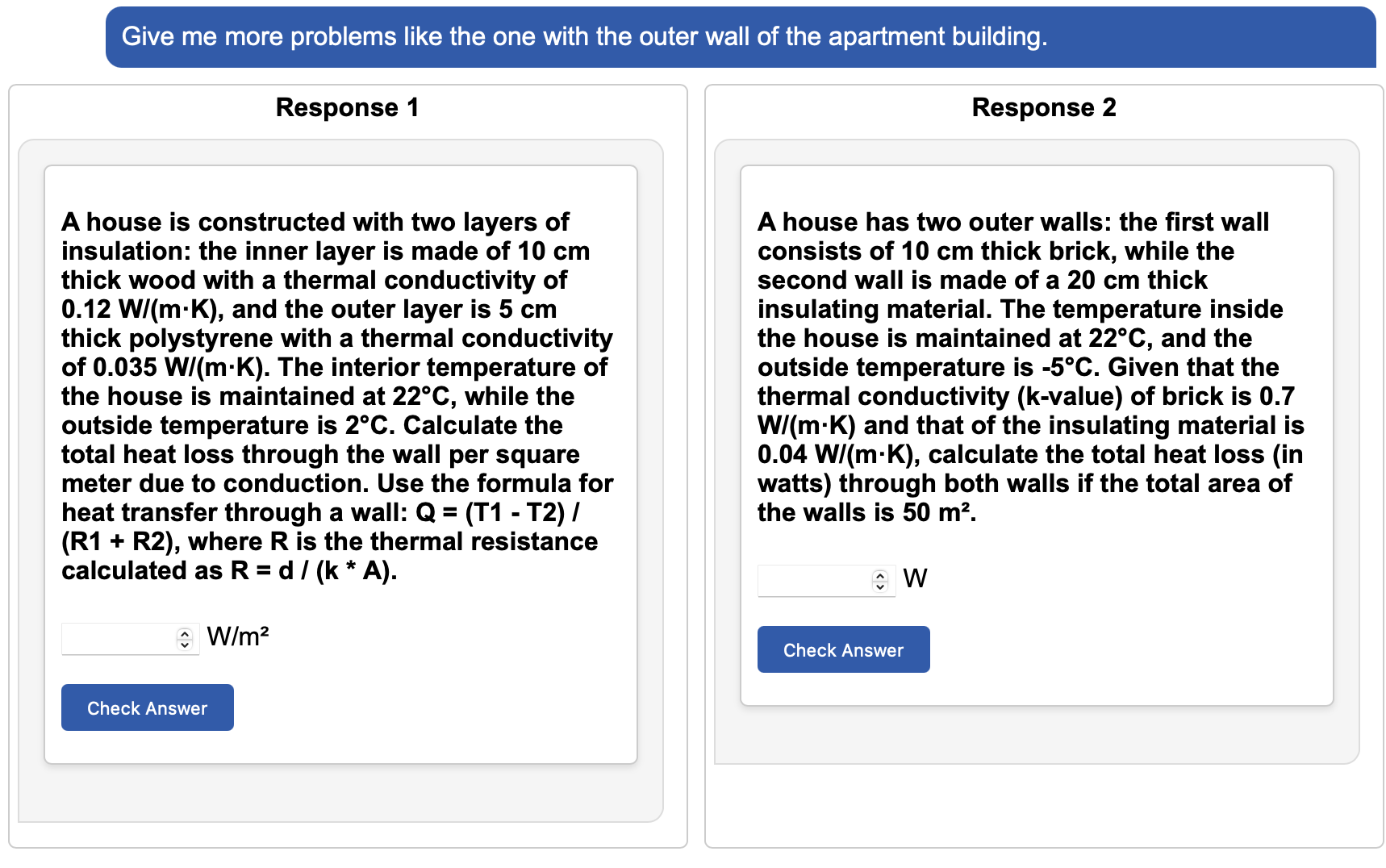}
\end{center}
\caption{Two interactive problems generated on-demand, showcasing how course-specific contextual information, in this case the premise of  the problem ``with the outer wall of the apartment building,'' is incorporated via Retrieval Augmented Generation (RAG)~\cite{Lewis2020Retrieval}. As opposed to traditional chatbot output as in Fig.~\ref{fig:example}, the problems are rendered with an interactive answer field.}
\label{fig:solve}
\end{figure*}

\begin{figure*}
\begin{center}
\includegraphics[width=0.8\textwidth]{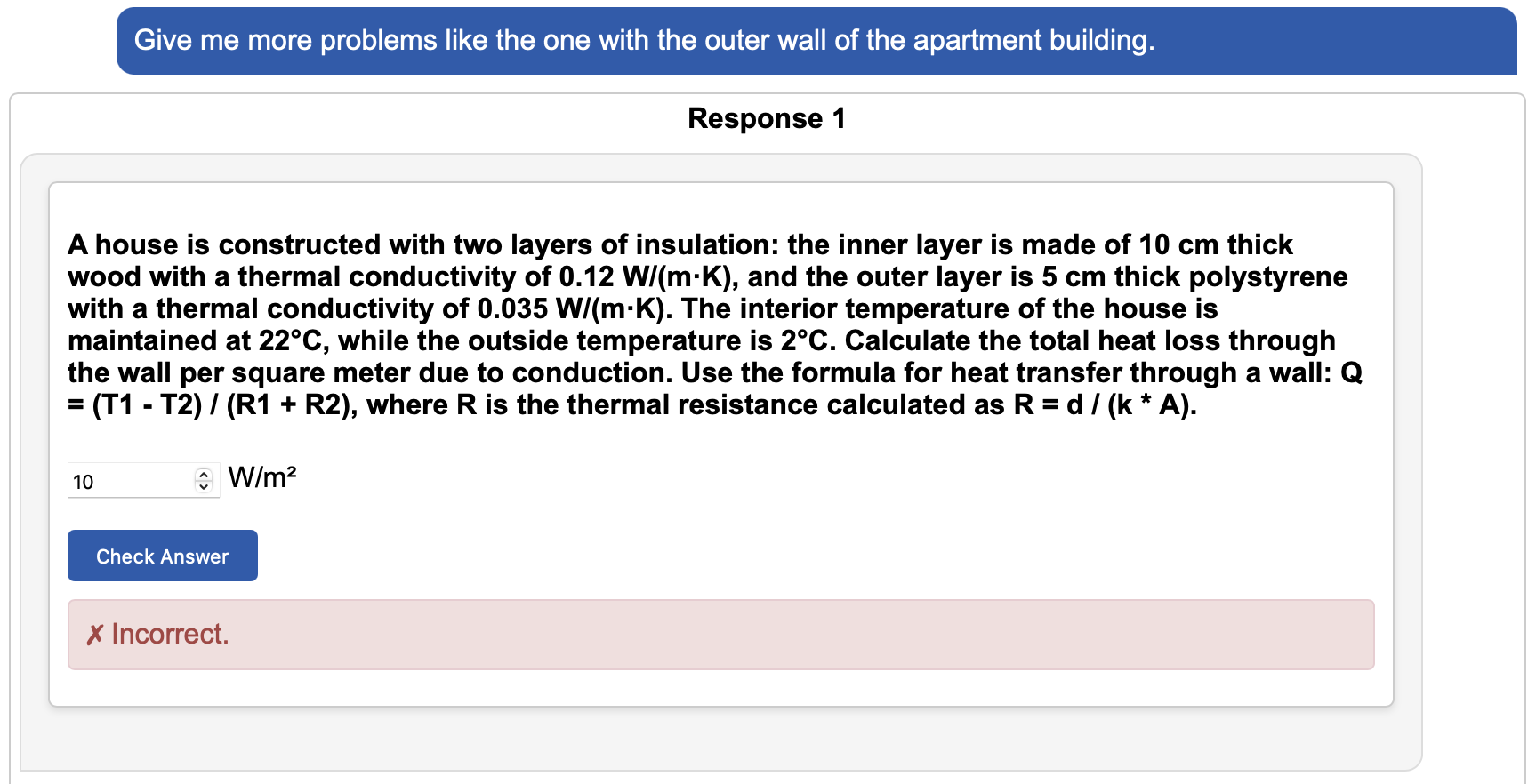}
\end{center}
\caption{Attempting to solve the selected problem from Fig.~\ref{fig:solve}.}
\label{fig:solveattempt}
\end{figure*}

To enable students to generate valid problems on demand from course materials, we explore automated metrics using LLM-as-a-judge methodologies~\cite{Gu2024Survey}. In contrast to prior studies that relied on synthetic data, we specifically evaluate problems produced during real interactions between students and an LLM-powered chatbot. Figures~\ref{fig:solve} and \ref{fig:solveattempt} show examples,  which also highlight how contextual information can effectively be integrated via Retrieval-Augmented Generation (RAG) techniques~\cite{Lewis2020Retrieval}.

\subsection{Formative assessment quality}\label{sec:qualities}
Formative assessment problems for physics are most effective when carefully designed to target key concepts and common difficulties. In this subsection, we use \emph{quality} in the pedagogical sense (supporting valid measurement and learning); later we separately operationalize \emph{relevance} as alignment with learner uptake in the deployment setting.
The quality of each problem, namely the question and its answer choices, is crucial for measuring understanding and promoting learning~\cite{scott2006evaluating}. As it turns out, generation of problems is one of the most frequent uses of AI by instructors~\cite{wattanakasiwich2025physics}.
In the context of independent study or small-group exam preparation, well-crafted problems (including solution paths~\cite{fakcharoenphol2011students}) can help students identify their own ideas about physics, practice problem-solving skills, and relate physics to the real world~\cite{gautreau1997concepts,fakcharoenphol2014physics,rodriguez2021frequent}. 

In multiple-choice problems, the incorrect options (``distractors'') should be written to be both plausible and diagnostic. Instead of obvious throw-away wrong answers, effective distractors often reflect common alternative or intuitive conceptions that students hold about the topic~\cite{gierl2017developing}. For example, a problem on Newton's laws might include a distractor aligning with the intuitive conception that a force is needed to keep an object moving~\cite{scott2018central}. Using known learner ideas as distractors serves two purposes: (1) it makes the problem challenging and discriminating (students who have the respective alternative conception will be tempted by that option, whereas those with a firm understanding will not), and (2) it provides learners with insight into their own thinking.

Good quantitative problems pay attention not only to concepts but also to the numerical values chosen for physical quantities. Research-based guidelines suggest using meaningful, realistic numbers that reinforce physical intuition and avoid unintended simplifications. If the context permits, values can be chosen to reflect real-world scales (for instance, using the mass of a car $\approx1000\mbox{kg}$ or a person $\approx70\mbox{kg}$, rather than overly simplified or non-physical values). Overall, choosing appropriate numerical values makes problems more authentic and educative: students must interpret results in context (e.g. recognize when an answer like $800\mbox{m/s}$ for a car's speed is unphysical) and engage in quantitative sense-making, rather than just number-crunching~\cite{heller1992teaching,mota2019homework} --- an important scientific skill.

Context-rich physics problems embed familiar, realistic scenarios, such as driving a car or playing sports, to tap into students' prior experiences, intuition~\cite{heller1992teaching}, or interests to increase motivation~\cite{walkington2019personalizing}. Unlike abstract problems, these problems often include extraneous or irrelevant data, requiring students to discern what information is actually needed. This design mirrors real-world situations, discourages rote formula use, and encourages deeper processing: students must translate the scenario into a solvable physics problem, fostering transferable problem-solving skills. Such richly contextualized problems may have the potential to improve engagement, comprehension, and motivation by making physics feel more meaningful and relevant~\cite{heller1992teaching,taasoobshirazi2008review}. Although more challenging, they can promote discussion and collaborative strategy-building, as they are often designed to be too difficult to solve mechanically alone but manageable in small groups, thus reflecting authentic scientific practice~\cite{heller1992teaching}. In this way, context-rich problems not only make learning more engaging but also develop students' ability to identify pertinent information and apply knowledge flexibly.

Each of the features above contributes to making formative assessment problems more effective learning tools. Targeted distractors ensure that a multiple-choice problem probes the student's conceptual understanding, rather than factual recall, by eliciting and addressing known misconceptions. Thoughtful numerical values and problem settings prevent students from bypassing understanding through rote computation; instead, they promote estimation, unit sense, and recognition of what a reasonable answer should be. Authentic contexts increase engagement and help students connect physics concepts to the world around them, which can improve knowledge retention and the ability to transfer learning to new situations. Formative assessment is most beneficial when it actively involves students in thinking like experts: checking their reasoning, confronting misconceptions, and making sense of results~\cite{dulger2025students}.

\subsection{Automated problem generation}
Historically, generating situational formative assessments on-the-fly and on-demand required experienced human tutors or instructors~\cite{harrison2018assessment}. Automated problem generation relied on rule-based algorithms, predefined templates, and ontology-driven methods, restricting flexibility~\cite{sadigh2012automating,Demaidi2017Evaluating,Nentwich2016Computer}. Natural-language processing (NLP)-driven approaches, such as those designed  to teach languages~\cite{AldabeArikIturri}, and transformer-based systems, such as those designed to generate programming exercises~\cite{FreitasNLP}, improved flexibility but remained limited by rigid structural dependencies.

Recent advancements in large language models (LLMs) have opened new possibilities for automated generation and evaluation of problems. However, current approaches often lack robust automated frameworks to filter out low-quality problems~\cite{kuchemann23}, use synthetic datasets insufficiently capturing authentic interactions~\cite{Chen2018LearningQ,Cobbe2021Training,Hendrycks2021MeasuringMP,Macina2023MathDial}, and rely on limited evaluation criteria weakly correlated with expert assessments~\cite{kortemeyer23ai,Doughty2023Comparative}. To address these gaps, our study aims to work with authentic and human-verified data: a comprehensive dataset of genuine student-chatbot interactions in an exam preparation context, capturing authentic student behavior, and a systematic evaluation of automated metrics against expert and student judgments, highlighting solution correctness, clarity, and cognitive complexity as key quality predictors. These contributions provide a foundation for improved reliability and effectiveness in AI-assisted education.

A core characteristic of generative AI is that it always produces \textit{some} output—regardless of its relevance, accuracy, or grounding. As a result, LLM-generated problems may be unclear, misleading, or entirely fabricated (``hallucinations''). Ensuring reliable, high quality automated problem generation, therefore, requires addressing both ends of the process:

\begin{enumerate}
\item Improving the output of generative AI systems~\cite{kortemeyer2024grading,chen2025grading}
\item Improving the evaluation of the quality of the output~\cite{kortemeyer2025assessing}
\end{enumerate}

The former is akin to an instructor drafting problems for an assignment or an exam and represents a broad research area; the latter parallels the ``proofreading'' done by colleagues or teaching assistants. In this study, we focus exclusively on the latter.

\subsection{Study objective}
The goal of our study is to identify validation metrics for problems generated by generative AI that are reliable, relevant, and automatically assessable.
 We employ an LLM-as-a-judge mechanism, where the initial LLM output is assessed by the LLM using a set of metrics. The goal of this validation is to distinguish viable problems that can be forwarded to the learners from defective problems that should be discarded.

We are using a dataset of problems generated on-demand by our LLM-powered chatbot during student-chatbot interactions in a user study emulating exam preparation. The system generated a problem whenever it detected that a student was requesting one; see Figure~\ref{fig:solve} for an example prompt. We establish ground truth on a set of potentially useful and usable quality metrics through an expert rating process; in other words, we consider these human expert ratings to be the ``truth'' on which we base the remainder of the study. We then establish the reliability of automatically assessing these same metrics using an LLM-as-a-judge, and we establish relevance by correlating these quality metrics with measures of user preferences.

\subsection{Operational definitions and research questions}\label{sec:rq}
We distinguish between pedagogical quality (an item's potential to diagnose understanding and support learning; see Sect.~\ref{sec:qualities}), technical defects, and relevance to learner uptake. While ideal validation would relate item properties to independent efficacy measures such as IRT-based item parameters or observed learning gains, those require repeated administrations or longitudinal designs that are outside the scope of this study and not available in our on-demand, conversational practice context. We therefore operationalize quality-related constructs within the constraints of our data and deployment setting.

Within the framework of our study, the terms \emph{reliable} and \emph{automatically assessable} refer to properties of the \emph{evaluation procedure}, while \emph{relevance} refers to observable student choice (not necessarily learning efficacy):
\begin{description}
\item[Reliable] Defined as model--ground truth agreement.
A metric is called \emph{reliable} if an LLM-as-a-judge can reproduce the expert label with high agreement.

\item[Relevant] Defined as predictive of learner choice and preference.
A metric is called \emph{relevant} if it is associated with what students actually prefer when selecting between candidate items.
Our system use case is that students are shown multiple candidate AI-generated problems and must choose what to attempt; therefore, a necessary condition for any pedagogically strong item to matter in practice is that it is actually selected and engaged with. We accordingly interpret relevance as perceived usefulness/appeal and practical fit (triangulated with exit-survey feedback), rather than as a direct measure of learning efficacy.

\item[Automatically assessable] Defined as feasible at scale without human intervention.
A metric is \emph{automatically assessable} if it can be computed end-to-end by an LLM from the generated problem (JSON) and its immediate chat context, returning a schema-conformant JSON judgment with no human intervention, external solution key, or additional retrieval beyond the conversation.
\end{description}

These definitions lead to the following research questions:

\begin{description}
  \item[RQ1 (Reliability)] Which quality metrics can commodity LLMs judge with strong agreement to expert labels on confirmatory samples, and how does this depend on model choice?
  \item[RQ2 (Relevance)] Which metrics are predictive of student choice between candidate problems for (a) multiple-choice and (b) numerical items, and how do these findings triangulate with free-form exit-survey feedback?
  \item[RQ3 (Automatic assessability)] Within the intersection of reliable and relevant metrics, which subset can be automatically assessed by an LLM fast and cheaply enough for real-time deployment, and how small can this subset be without materially degrading predictive power?
\end{description}

\section{Methodology}
The following subsections describe the implementation of our study. Note that in several of the technical implementation details, for example the labels of the metrics and the prompts, practice problems were denoted as exercises and in rare cases as questions --- we have reproduced these in the original.

\subsection{Target system}
\begin{figure*}
\begin{center}
\includegraphics[width=0.8\textwidth]{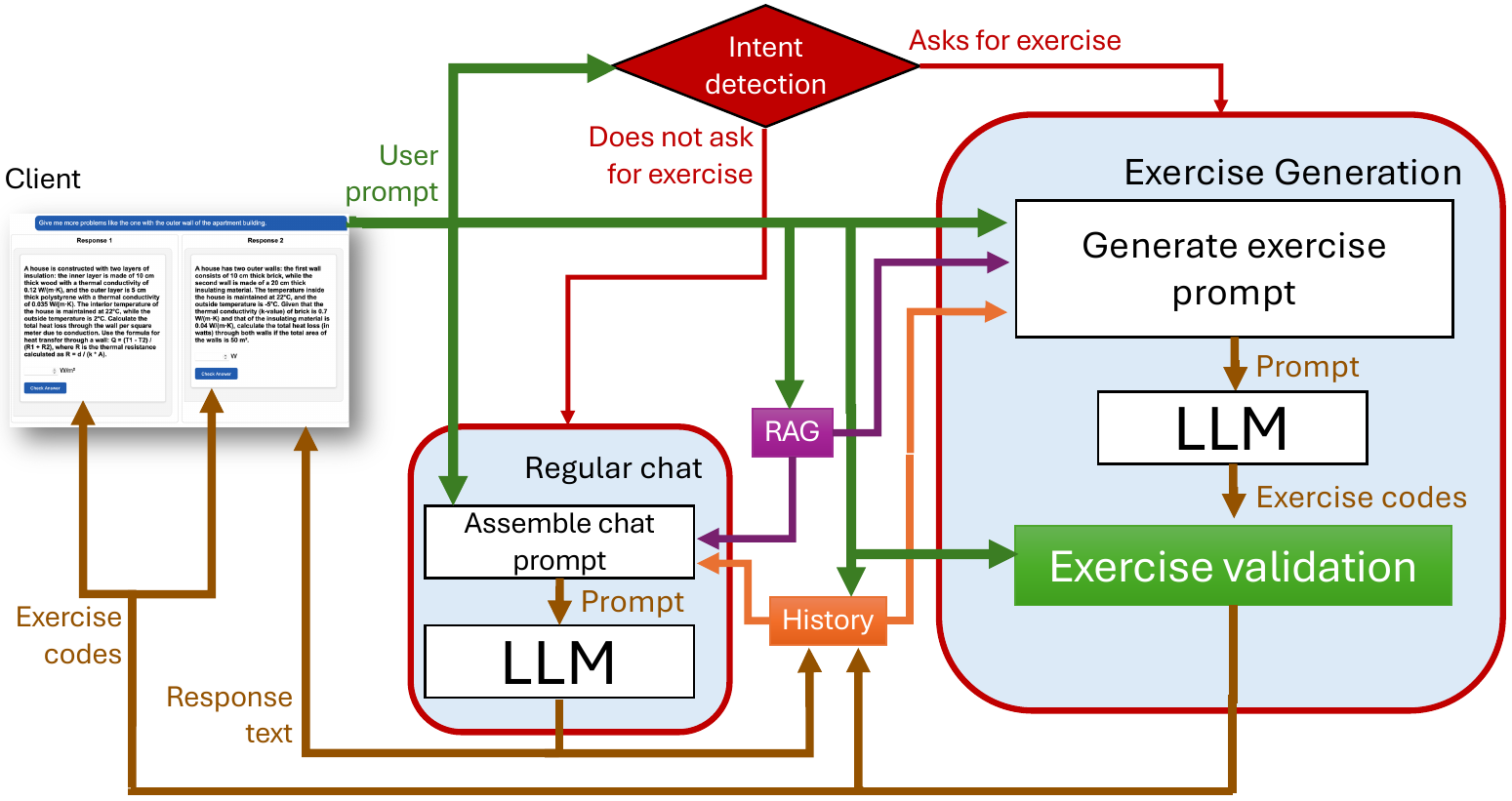}
\end{center}
\caption{Information flow in an enhanced chatbot that can generate verified practice problems on-the-fly and on-demand. The ``problem validation'' step (LLM-as-a-judge; green box) is the subject of our study.}
\label{fig:target}
\end{figure*}

The eventual goal of our research and development effort is to enhance current chatbots with the ability to generate verified problems on-the-fly and on-demand; Fig.~\ref{fig:target} gives an overview of the information flow in such a system. The technology platform for our study is provided by Ethel~\cite{kortemeyer2024ethel}, a virtual teaching-assistant ecosystem developed at ETH~Zurich.

When a user submits a new prompt in the chat interface, the system first determines intent: does the user request an interactive practice problem or a regular conversational response? This intent detection is performed by a smaller ``mini'' LLM. If the user does not appear to request a practice problem, the standard chat mechanism of Ethel generates the response. This standard mechanism leverages RAG to incorporate relevant information from course materials and utilizes chat-history memory to maintain conversational continuity.

If the user's intent indicates a desire for an interactive practice problem, the system formulates a specialized prompt using the same foundational information as in regular chat, augmented with additional instructions that guide the LLM to produce structured JavaScript Object Notation (JSON) outputs adhering to predefined schemas for various interactive problem formats~\cite{liu2024we}. Preliminary research indicated that the quality of generated problems varies widely; thus, multiple candidate problems are simultaneously generated. The subsequent validation of these candidate problems, employing automated quality metrics, forms the core focus of our current research.

Once validated, the JSON data representing selected problems is transmitted to the user's browser, where JavaScript dynamically renders interactive problems based on this structured data. Client-side scripts evaluate student-submitted answers for correctness (see Fig.~\ref{fig:solveattempt}). Students can continue their dialogue with the chatbot or revisit and retry previously generated problems by scrolling through the chat history.

During the dataset generation for this study, only $1$-out-of-$N$ multiple-choice (``radio-button'') and numerical problem types were implemented. Problem validation at this stage involved only basic sanity checks, resulting in a dataset with varying and frequently insufficient quality.

\subsection{Dataset}
We recruited 34~students from an introductory physics course at ETH~Zurich to prepare for an upcoming exam in a controlled laboratory setting during one of three sessions; the exam covered both semesters of the course. During the running semester, participants, like all students in the course, already had unlimited access to the standard conversational version of  Ethel; for the purpose of this study, participants were given 1.5~hours of access to a  version that was enhanced according to Fig.~\ref{fig:target}.

They were asked to use the time in the laboratory to prepare for the following topics that were going to be covered on their upcoming exams:
\begin{itemize}
\item Mechanical energy and power
\item Kirchhoff rules in DC~circuits
\item Electrical field and potential
\end{itemize}

Typical examples of student prompts which triggered problem generation were, translated from German, are:
\begin{itemize}
\item Please give me an entry-level problem about torque.
\item Give me an exam problem about mechanical energy.
\item Please give me a problem involving Kirchhoff rules.
\item Give me a conceptual question about kinematics.
\item Give me complex problem about electric fields.
\item Give me a more difficult problem about kinematics.
\item Another one.
\end{itemize}
For prompts like the last one, ``another one,'' the intent detection concluded that the user desired another problem like the most recently generated one. Generally, problem prompts were not very specific when it came to desired scenarios or solution methods.
Examples of general prompts, again translated from German, are:
\begin{itemize}
\item How do I find the sign of the torque?
\item Please explain the solution path for this problem.
\item What are 0.08 mJ in J?
\item Why is this formula only valid for a straight path?
\end{itemize}

A total of $543$~problems were generated in response to the participants, where one student prompt could result in more than one problem (see Fig.~\ref{fig:solve}). Students would select the problem candidate that they wanted to work with and attempt to solve. For performance reasons, in later laboratory sessions, the answer for numerical problems would not be generated by the server until after the student selected the candidate. For each generated problem, the following data were collected:
\begin{itemize}
\item Preceding prompts and responses
\item JSON code of the problem
\item Any solution attempts
\item If a solution had been generated
\end{itemize}

Additional data collected included a post-survey on their experiences, which allowed for free-form entry.  It asks about perceived usefulness and whether the generated problems met expectations, as well as the quality, clarity, and trustworthiness of the chatbot's answers and explanations. It also probes perceived learning efficiency compared to traditional methods, effects on motivation, and how well the chatbot adapted to individual learning needs. The Likert-type questions (translated from German to English) were:
\begin{itemize}
  \item How useful was the chatbot for your preparation for the mini-exam?
  \item How useful did you find the automatic generation of practice problems?
  \item Did the automatic generation of practice problems meet your expectations?
  \item How do you rate the quality of the answers generated by the chatbot?
  \item How clear were the chatbot's explanations?
  \item How trustworthy did you find the chatbot's information?
  \item How efficient was learning with the chatbot compared to traditional methods?
  \item How did using the chatbot affect your motivation to learn?
  \item How well was the chatbot able to address your individual learning needs?
\end{itemize}
The survey also included an open-ended field for comments, remarks, etc.

\subsection{A note about nomenclature}
In this manuscript, we use \textit{problem} to mean a typical exam or homework item consisting of a \textit{question} prompt and an associated \textit{answer} (e.g., a solution or answer choices). In the analysis code and data structures created earlier in the project (repository listed under Data Availability), the terms \textit{problem} , \textit{task}, and \textit{exercise} appear as legacy variable names and are used interchangeably; for transparency and reproducibility, we retain those identifiers in the shared code and tables provided here, but we use the standardized terminology in the remainder of the manuscript. Finally, when students used the German term \textit{Fragen}, we translate it as \textit{questions}, even when the context indicates they were referring to complete \textit{problems} as defined above.

\subsection{Validity metrics}
The metrics we used for validating the generated problems were inspired by work on metrics for problem solutions~\cite{docktor2016assessing}; we asked ourselves what properties a problem would need to have to demand solutions that address those solution metrics. We added metrics that assess the implicit quality criteria of solvability, completeness, and correctness of the expected answer. Both sets of metrics align with earlier work on assessing the quality of student-generated problems~\cite{bates2014assessing}. To our knowledge, the resulting set is the largest set assembled thus far in literature, however, we still cannot claim completeness; we thus designed the validation mechanism for possible expansion in future studies.

Table~\ref{tab:manual} shows quality metrics that are  determined from the problem and the preceding dialogue. Table~\ref{tab:both} shows metrics that are determined based solely on the generated problem for both problem types; Table~\ref{tab:num} shows additional metrics that only pertain to numerical problems, while Table~\ref{tab:mc} shows the same for $1$-out-of-$N$ multiple-choice problems. Here, ``N/A'' is an explicit ground-truth label indicating that the desired level was not specified or inferable from the chat context (rather than an unscored item). For the agreement analyses in Sect.~\ref{sec:perfllm} we treat N/A as an additional class for these intent metrics; their low macro-level scores therefore reflect that the intended Bloom/difficulty level is often not recoverable from the available context. For the relevance models in Sect.~\ref{sec:choices}, we excluded these two predictors (rather than excluding the corresponding items), so the remaining analyses are unaffected except that we do not make claims about the influence of these intent-level constructs on student choice.

\begin{table*}
  \renewcommand{\arraystretch}{1.4}
\caption{Metrics derived from preceding prompts as part of the ground truth. In our technical implementation, the terms \textit{exercise} and \textit{(practice) problem} were used interchangeably.\label{tab:manual}}
\begin{ruledtabular}
% --- Table for ``Metrics for Both Answer Types'' ---
\begin{tabular}{p{4.5cm} p{5.5cm} p{6.5cm}}

{Identifier} & {Definition and Context} & {Evaluation and Rating Criteria} \\ \hline
\textit{expected-bloom-level} & Indicates the Bloom level (1--4 or \textit{N/A}.) the user presumably wants (or that is implied in conversation). The highest level, ``create,'' does not apply in our introductory physics context. If undetermined, it is \textit{N/A}. & 1:~Remember \newline2:~Understand \newline3:~Apply \newline4:~Analyze\newline\textit{N/A}:~No clear assignment can be made based on the chat history.\\

\textit{expected-student-difficulty-level} & Reflects the difficulty level (0--4 or \textit{N/A}) that the user explicitly or implicitly requests. If no statement is made, defaults to \textit{N/A}.&0:~Intuitive (common-sense) \newline1:~Foundational (straightforward) \newline2:~Intermediate (multi-step, multiple concepts; mid-quartiles) \newline3:~Advanced (complex; upper quartile) \newline4:~Expert (beyond student level)\newline\textit{N/A}:~No clear assignment can be made based on the chat history.\\

\textit{exercise-is-relevant-to-user-request} & Measures how well the problem aligns with the user's explicitly requested topic or learning objective. & 2:~The user's requested topic is central. \newline1:~Some partial match, but not the main focus. \newline0:~Irrelevant or off-topic.
\end{tabular}
\end{ruledtabular}
\end{table*}

\begin{table*}
  \renewcommand{\arraystretch}{1.4}
\caption{Metrics used for both numerical and 1-out-of-$N$ (``radio-button'') problems. In our technical implementation, the terms \textit{exercise}, \textit{task}, and \textit{(practice) problem} were used interchangeably. \label{tab:both}}
\begin{ruledtabular}
\begin{tabular}{p{4.5cm} p{5.5cm} p{6.5cm}}

{Identifier} & {Definition and Context} & {Evaluation and Rating Criteria} \\ \hline

\textit{bias-free-language} & Language used in the problem is neutral and respectful, free of stereotyping or discriminatory content. & 1:~No biases or discriminatory remarks are present. \newline0:~Biased or offensive language.\\

\textit{bloom-level-of-exercise} & Classifies the cognitive demand according to Bloom's Revised Taxonomy. The highest level, ``create,'' does not apply in our introductory physics context. & 1:~Remember \newline2:~Understand \newline3:~Apply \newline4:~Analyze\\

\textit{contains-harmless-extra-info} & The problem includes superfluous details or narrative elements that do not confuse or contradict the solution. & 1:~Extra information is present but not contradictory or confusing. \newline0:~No or misleading extra info.\\

\textit{contains-misleading-extra-info} & Flags contradictory or irrelevant data that confuses the solver or renders the problem unsolvable/ambiguous. & 1:~Contradictory or confusing data is present. \newline0:~No such misleading information.\\

\textit{contains-relatable-extra-info} & Indicates whether the problem's additional details connect to students' everyday experiences or  real-world contexts that promote engagement. & 1:~Extra info clearly mirrors common or familiar situations. \newline0:~No real-life framing, or purely abstract.\\

\textit{exercise-difficulty-level} & Rates how difficult the problem is for a typical undergraduate physics student, from 0 (intuitive) to 4 (expert-level challenge). & 0:~Intuitive (common-sense) \newline1:~Foundational (straightforward) \newline2:~Intermediate (multi-step, multiple concepts; mid-quartiles) \newline3:~Advanced (complex; upper quartile) \newline4:~Expert (beyond student level)\\

\textit{grounded-in-textbook} & The problem aligns with the course materials.  & 1:~Matches standard course topics. \newline0:~Off-curriculum or too advanced for typical undergrad references.\\

\textit{includes-solution-strategy} & The problem provides a hint (``Use conservation of energy \ldots''), without revealing the solution. & 1:~A succinct strategic hint is given. \newline0:~No hint, or the solution steps are fully spelled out in the prompt.\\

\textit{llm-solution-is-correct} & Evaluates whether the final answer (if any) provided by the LLM against which student input will be checked is correct. 
 & 1:~The final answer is correct. \newline0:~Answer is inaccurate or flawed. \\

\textit{realistic-data-used} & Evaluates whether numeric values are physically reasonable. & 1:~Values lie within plausible ranges. \newline0:~Data is absurd or contradictory (e.g., a 200\,kg fly).\\

\textit{solution-not-in-problem} & The problem refrains from providing the final answer. Hints may appear, but not the full solution path or result. & 1:~No final answer  is revealed in the statement. \newline0:~The final answer or a fully worked-out approach is revealed.\\

\textit{task-is-clear} & Indicates whether the problem statement is phrased unambiguously so that students know precisely what is being asked. & 1:~The instructions are direct, variables clearly defined, and no confusion about what to solve. \newline0:~Missing or ambiguous instructions leave the student uncertain.\\

\textit{task-is-specific-and-complete} & Verifies that all essential information is present: variables, boundary conditions, units, assumptions. Nothing critical is omitted. & 1:~No key data is missing; the problem can be tackled directly. \newline0:~Required information is absent or unclear.\\

\textit{task-is-solvable}&A unique solution exists, and the problem includes sufficient, non-contradictory
information to arrive at it.&
1:~The problem can be solved.\newline
0:~The problem is unsolvable..
\\

\textit{uses-standard-notation-concepts}&
Uses standard terminology, concepts, notations, language and expressions.&
1:~accepted language, etc.\newline
0:~uses unusual, non-standard language, etc.
\end{tabular}
\end{ruledtabular}
\end{table*}

\begin{table*}
  \renewcommand{\arraystretch}{1.4}
\caption{Additional metrics used for numerical problems.\label{tab:num}}
\begin{ruledtabular}
\begin{tabular}{p{4.5cm} p{5.5cm} p{6.5cm}}

{Identifier} & {Definition and Context} & {Evaluation and Rating Criteria} \\ \hline
\textit{asks-for-exactly-one-solution} & Determines if the problem is structured to yield a single valid solution under typical physics constraints. A problem with multiple correct answers or ambiguous conditions would not meet this criterion. & 1:~The problem setup ensures exactly one valid solution. \newline0:~Multiple or ambiguous solutions exist.\\

\textit{asks-for-numerical-answer-only} & Indicates whether the task specifically demands a purely numerical answer. Answers needing symbolic expressions, proofs, or multiple-choice selections do not qualify. & 1:~The problem clearly requests a single number (possibly with units). \newline0:~The solution format is not strictly numeric (may include conceptual explanations or multiple-choice).\\

\textit{measurement-unit-is-clearly-stated}&
Checks whether the problem explicitly states the desired unit for the answer (``in meters,'' ``in Joules,'' etc.).
&
1:~The problem explicitly mentions units for the final answer.\newline
0:~The problem does not specify units.
\end{tabular}
\end{ruledtabular}
\end{table*}

\begin{table*}
  \renewcommand{\arraystretch}{1.4}
\caption{Additional metrics used for 1-out-of-$N$ (``radio-button'') problems.\label{tab:mc}}
\begin{ruledtabular}
\begin{tabular}{p{4.5cm} p{5.5cm} p{6.5cm}}
{Identifier} & {Definition and Context} & {Evaluation and Rating Criteria} \\ \hline
\textit{distinct-misconceptions} & Ensures that each incorrect option in a multiple-choice problem represents a different common error or misunderstanding, rather than being a rephrasing of the same mistake. & 1:~Each incorrect option is uniquely wrong for a specific,  commonly occurring  reason. \newline 0:~Two or more distractors replicate the same fundamental mistake.\\

\textit{non-partial-options} & Ensures that each multiple-choice option is fully correct or fully incorrect, with no ambiguous half-truths. & 1:~No distractor is ``semi-correct.'' \newline0:~At least one distractor is partially correct, making it unclear.\\

\textit{single-correct-option} & Verifies that exactly one choice in a multiple-choice problem is correct. If the problem designates one right answer but all are actually incorrect, it can still be considered 1 for that design. & 1:~Exactly one choice is correct. \newline0:~More than one or none are valid answers in a supposed single-choice format.\\

\textit{uniform-option-style} & The options are of similar style, detail, and length, with the exception of standard options like ``none of the above.'' & 
1:~The options, both incorrect and correct ones, are uniform.\newline
0:~One or more options are jumping out by being dissimilar in nature.
\end{tabular}
\end{ruledtabular}
\end{table*}

Figure~\ref{fig:definition} shows an example of how these metrics were transcribed within our LLM-as-a-judge pipeline as a JSON structure. This description of the respective evaluation metric, the problem being evaluated, and the evaluation instructions are input into a prompt template to elicit a structured JSON response from the LLM, which the evaluation pipeline then processes.

\begin{figure*}
\scriptsize
\begin{verbatim}
    {
      "metric_id": "contains-misleading-extra-info",
      "human_metric": "Contains misleading extra information, which makes the exercise contradictory or unsolvable.",
      "definition": "Evaluates whether the problem contains extra information that contradicts or conflicts with the necessary data,
      rendering the exercise unsolvable or ambiguous.  Extra information is flagged as misleading only if it creates contradictions
      or imposes conflicting constraints without pedagogical value.",
      "rating": "1: This is the case\n0: Not the case.",
      "radio_only": false,
      "numerical_only": false,
      "short_title": "Contains Misleading Extra Info",
      "category": "Clarity and Specificity",
      "control_type": "checkbox",
      "editable": true,
      "visible": true,
      "rating_mapping": {
        "true": "this is the case",
        "false": "not the case"
      },
      "rating_clarifications_and_guidelines": "Extra information is considered misleading if it contradicts other provided data
      or adds conflicting constraints without offering pedagogical value. Information that is extra but aids in learning
      (e.g., irrelevant details that do not cause confusion) should not be marked as misleading.",
      "examples": "
      - Example: A problem that states conflicting values for a variable
      Expected Evaluation: 1\n
      - Example: A problem that includes an extra detail such as \"the ball is red\" which does not interfere with the solution 
      Expected Evaluation: 0",
      "dataset_properties_used": ["exercise"]
    }
\end{verbatim}
\caption{Example of the definition  for \textit{contains-misleading-extra-info} in our LLM-as-a-judge pipeline (green box in Fig.~\ref{fig:target}).}
\label{fig:definition}
\end{figure*}

\subsection{Ground truth}\label{sec:gt}
To establish the ground truth, the problems and dialogues were evaluated on the quality metrics by one of the authors (G.K.; T.G.~is a computer scientist), who taught introductory physics courses for over two decades. This manual rating of  $\approx10,000$~metrics took several days.

\subsubsection{Class imbalance (``skewedness'')}\label{sec:skew}
A \emph{class} $C_i$ denotes one of the discrete, mutually exclusive outcome labels that a given metric can take. 
For example, for \textit{task-is-solvable} the classes are $\{0,1\}$; for \textit{bloom-level-of-exercise} the classes are $\{1,2,3,4\}$; and for \textit{exercise-is-relevant-to-user-request} the classes are $\{0,1,2\}$.

Although we collected a total of $N = 543$ generated problems, the human-rated class distribution for some metrics was highly skewed --- indeed, certain classes $C_i$ were underrepresented or even absent.  To ensure that per-class estimates (e.g., precision or recall for each $C_i$) have sufficiently low sampling variability before treating them as definitive, we adopted the conventional rule of requiring at least 30 observations per class.  This ``30-sample rule'' is a rule of thumb based on the central limit theorem: binomial proportions begin to approximate the normal distribution when $n\gtrsim30$, yielding 95~\% confidence interval half-widths on the order of $\pm18$ percentage points or less~\cite{Hogg2019_PSI,Orawo2021_BinomialCI}. 

Accordingly, we label any metric for which every class count $n_{C_i}\ge30$ as confirmatory, indicating that standard inferential interpretations apply; metrics for which one or more classes fall below this threshold are regarded as exploratory, meaning they serve primarily to generate hypotheses rather than to support firm conclusions.  
Particularly the following metrics were unbalanced and thus only exploratory, but are of interest:

\begin{description}
    \item[\textit{bias-free-language}] In all 543~problems, there was no case of biased language or offensive content. Gender-neutral language was used in all cases. Subjectively, most problems were context-independent, focusing on circuit elements or blocks on slopes; these theoretical, ``clean room'' situations were not at risk of biases.
    \item[\textit{grounded-in-textbook}] There were only five cases of problems that were not grounded in the textbook, and those turned out to be hallucinations like a biology problem about mitosis or a problem about the seasons on Earth (``hallucinations'' in this context refers to instances where an LLM generates plausible-sounding but inaccurate or entirely fabricated information~\cite{ji2023survey}).
    \item[\textit{realistic-data-used}] There was only one problem that had unrealistic data, namely a problem with a freight train that has a total mass of 5,000 kg, i.e., five tons, which would be the mass of a small truck.
    \item[\textit{exercise-is-relevant-to-user-request}] In only seven cases was the generated problem not relevant to the user request, and among those were the five cases that were not grounded in the textbook.
\end{description}

\subsubsection{Interrelatedness}\label{sec:inter}
We now investigate how metrics in the ground truth are correlated with each other. Added here is the variable \textit{chosen} if the student chose to work on this problem in the forced-choice shown in Fig.~\ref{fig:solve}; this metric reflects student preferences and the problem's perceived alignment with desired features (one might argue that students may not know ``what's good for them,'' however, we are modeling a scenario where students are motivated by studying for an exam, not a homework scenario where they might choose the path of least resistance). This may allow us to later use a metric that is more reliable as a proxy for relevant, but less reliable, metrics.

\begin{figure}
\begin{center}
\includegraphics[width=1.0\columnwidth]{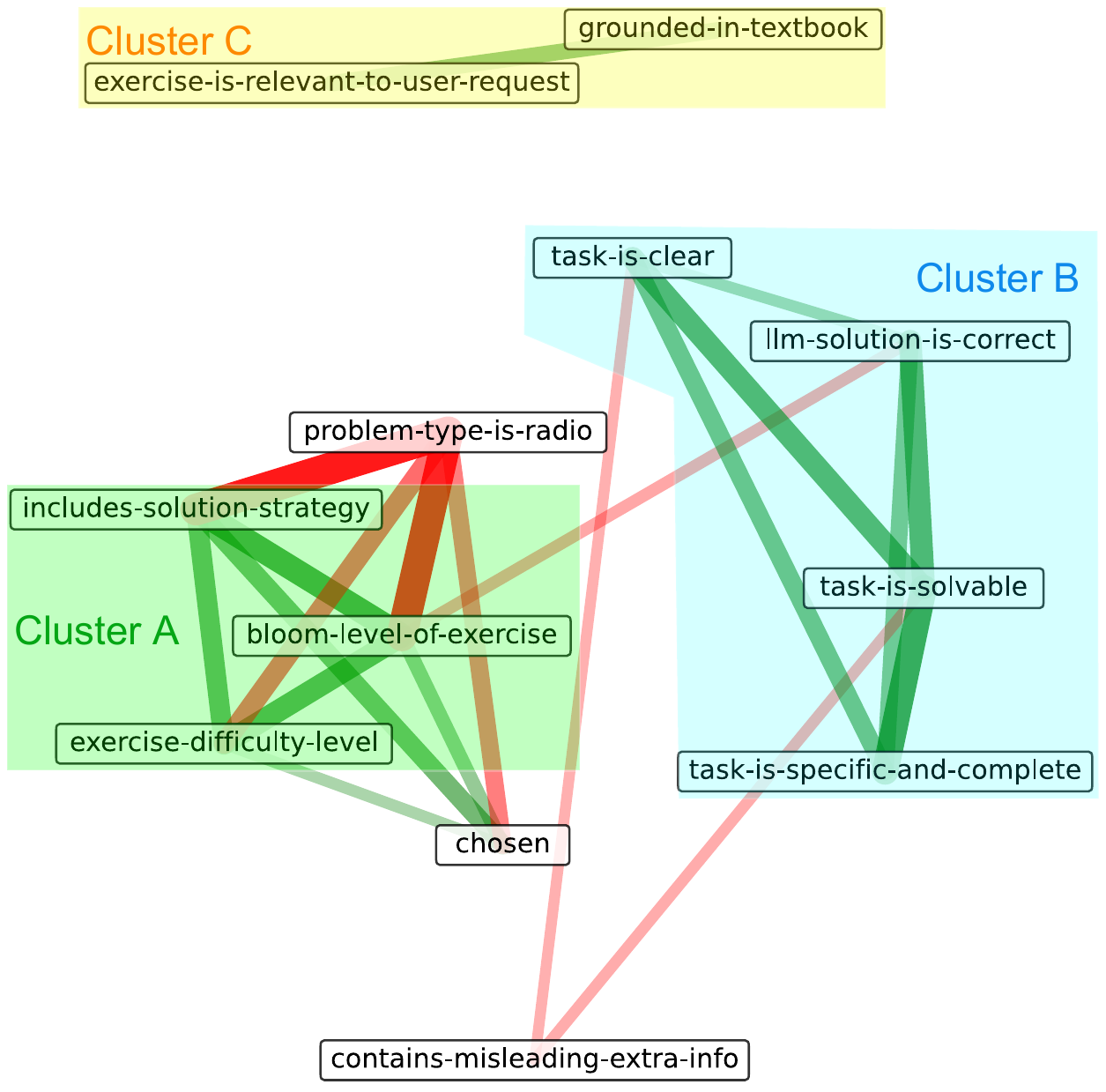}
\end{center}
\caption{Correlations between the metrics in a force-directed Fruchterman-Reingold representation~\cite{fruchterman1991}. Positive correlations are shown in green, negative correlations in red. The thickness of the lines indicates the Spearman correlation strength~\cite{spearman1904}. Any edges corresponding to an absolute correlation value $\varrho<0.2$ and isolated nodes have been removed. All shown correlations are highly significant (p $< 0.001$). The clusters identified in Sect.~\ref{sec:inter} are highlighted in green, blue, and yellow, respectively; the variables \textit{chosen} and \textit{problem-type-is-radio} are also connected to Cluster~A, but not determined by the problem-validation step in Fig.~\ref{fig:target}.}
\label{fig:correls}
\end{figure}

The force-directed Fruchterman--Reingold representation~\cite{fruchterman1991} in Fig.~\ref{fig:correls} places the metrics, represented as nodes, in two dimensions by minimizing the equivalent of potential energy where adjacent nodes attract (spring forces proportional to their pairwise Spearman correlations $\varrho$~\cite{spearman1904}), and all nodes mutually repel (see Ref.~\cite{kortemeyer2022virtual} for a physics derivation of this layout); the result is a geometry in which strongly connected variables appear closer together and clustering becomes visually apparent.

Three clusters emerge from Fig.~\ref{fig:correls}, based on pairwise highly significant Spearman correlations $\varrho\ge0.2$ (p $< 0.001$) :

\begin{description}
  \item[Cluster A -- Conceptual Depth]  
        \textit{bloom-level-of-exercise},
        \textit{exercise-difficulty-level},
        and \textit{includes-solution-strategy} form a tight triangle
        with $\rho\ge0.45$.
        The node \textit{chosen} connects positively to all three
        (e.g.\ $\rho=0.43$ with \textit{includes-solution-strategy}),
        indicating that students are attracted to tasks they perceive
        as cognitively engaging \emph{and} well scaffolded. It is surprising to find the close association between \textit{includes-solution-strategy} (a purely structural feature) and depth measures of difficulty and Bloom level.
  \item[Cluster B -- Task Clarity \& Feasibility]  
        \textit{task-is-clear},
        \textit{task-is-solvable},
        \textit{task-is-specific-and-complete}, and
        \textit{llm-solution-is-correct} form a second group on the
        right-hand side of the graph.
        Internal correlations range from $\rho=0.26$ to $0.59$,
        suggesting a shared latent trait of ``problem description quality.''
        In particular,
        \textit{llm-solution-is-correct} is strongly aligned with
        \textit{task-is-solvable} ($\rho=0.51$) and
        \textit{task-is-specific-and-complete} ($\rho=0.39$) but
        \emph{negatively} linked to Bloom level
        ($\rho=-0.27$). 
  \item[Cluster C -- Grounding \& Relevance]  
        A smaller pair connects
        \textit{exercise-is-relevant-to-user-request} with
        \textit{grounded-in-textbook} ($\rho=0.36$).
        Because both metrics show minimal variance, their practical impact
        on choice is limited, and they remain peripheral in the graph.
\end{description}

A salient node that ties together other nodes (a ``hub'') is \textit{problem-type-is-radio}, which
correlates inversely with nearly every node in Cluster A (``negative hub''), most
pronounced with \textit{includes-solution-strategy}
($\rho=-0.71$) and Bloom level ($\rho=-0.68$).
This mirrors the feature-importance result that radio items are less
likely to be selected when a more structured, quantitatively framed
alternative is available.

\subsubsection{Subjective observations}
While establishing the ground truth, all problems generated in this study had to be read and solved.
Subjectively, they contained a disproportionate number of run-of-the-mill standard problems, which were generally solvable in one or two steps. The problems generally lacked sophistication or interest, similar to typical end-of-chapter problems in non-research-based textbooks. The scenarios and tasks were at times repetitive with only slight variations, and they were generally not situated in authentic contexts.

\subsection{Considered LLMs}\label{sec:llms}
Generally, large language models are classified as ``reasoning'' or ``non-reasoning.'' Reasoning LLMs are specifically designed and fine-tuned to perform step-by-step logical, mathematical, or causal reasoning beyond surface-level pattern matching, whereas non-reasoning LLMs generate responses primarily by statistically mimicking observed text patterns without explicit internal reasoning processes. This distinction reflects whether the model actively constructs intermediate inferential steps or merely retrieves and recombines plausible sequences from its training data. Reasoning models are generally slower and more expensive to use.

Due to the high computational effort involved in evaluating the metrics in Tables~\ref{tab:manual}--\ref{tab:mc}, we limited ourselves to the lower-end, more affordable models offered at the time by OpenAI~\cite{openai} through Azure~\cite{azure}, namely:
\begin{description}
\item[GPT-4o] general purpose, non-reasoning model~\cite{gpt4o}
\item[GPT-4o-mini] limited, low-latency version of the non-reasoning GPT-4o~\cite{gpt4omini}
\item[GPT-o3-mini/low]  limited, low-latency version of reasoning model GPT-o3, set to low reasoning depth~\cite{gpto3mini}
\end{description}
Because the study ran on the shared Ethel platform, whose resources simultaneously served 3000 students across 20 courses at that time, we were unable to separate its actual computational cost from general operations; we are thus unable to assign a price-tag to the mechanisms under investigation.

\subsection{Measures of metrics judgment performance}
Several of our metrics have multiple classes $C_i$; $i=1\ldots n$, where $n\ge2$. An example is \textit{bloom-level-of-exercise}, where $i$ runs from $1$ to $n=4$. The traditional measures for agreement of the model's judgment and the ground truth are then defined on a per-class basis as
\begin{description}
\item[True Positive ($\mbox{TP}_i$)] The number of examples that truly belong to class  $C_i$ and were judged as $C_i$.
\item[False Positive ($\mbox{FP}_i$)] The number of examples that truly belong to some class other than $C_i$ (e.g., $C_k$, $k\ne i$)  but were judged as $C_i$.
\item[False Negative ($\mbox{FN}_i$)]  The number of examples that truly belong to $C_i$ but were judged as belonging to some other class $C_k$, $k\ne i$.
\item[True Negative ($\mbox{TN}_i$)]  The number of examples that neither belong to $C_i$ nor were judged as belonging to $C_i$ (i.e., everything else that's correctly classified as not $C_i$).
\end{description}
When comparing the judged values of the metrics to the ground truth, we took into account the unbalanced nature of our dataset by considering macro-level scores, which are the arithmetic mean of the individual class scores:
\begin{eqnarray}
\text{Accuracy} &=& \frac{1}{n}
\sum_{i=1}^{n}\frac{\text{TP}_i + \text{TN}_i}{\text{Number of Samples}}\\
\text{Precision}& =&\frac{1}{n}
\sum_{i=1}^{n} \frac{\text{TP}_i}{\text{TP}_i + \text{FP}_i}\\
\text{Recall}&=&\frac{1}{n}
\sum_{i=1}^{n} \frac{\text{TP}_i}{\text{TP}_i + \text{FN}_i}\\
\text{F1}&=&\frac{1}{n}\sum_{i=1}^{n}
  \frac{2\,\text{TP}_i}{2\,\text{TP}_i + \text{FP}_i + \text{FN}_i}
\end{eqnarray}

\subsection{Measures of user choices}
In the user study, when students requested a problem from the chatbot, they were presented with two alternative problems to choose from (see Figure~\ref{fig:solve}); this choice is captured by the metric \textit{chosen}. In addition to  Spearman correlations~\cite{spearman1904} of metrics to \textit{chosen}, we also analyzed which metrics predict \textit{chosen}. We employed random forest models (a type of ensemble machine learning algorithm)~\cite{breiman2001random}, which we trained for each problem type ($1$-out-of-$N$ (radio-button) and numerical problems).

We split the dataset $80/20$ into training and held-out test sets. Within the 80\% training portion we performed $5$-fold cross-validation to tune the Random Forest's hyperparameters, and then applied the final model to the remaining 20\% test set for evaluation. To work with sufficient sample sizes, we dropped features with  $>30\%$ missing or non-applicable values (``N/A'').

The performance of these models was evaluated using two measures: training accuracy, reflecting how well the model fits the training data, and the out-of-bag (OOB) score, an internal validation metric specific to random forests. The OOB score provides an unbiased estimate of a model's ability to generalize to unseen data, making it particularly useful for assessing predictive reliability. Additionally, feature importance scores were extracted from each model to identify and rank which problem characteristics most strongly influenced student decisions.

\subsection{Thresholds}
We are setting the following thresholds for our three quality goals:
\begin{description}
\item[Reliable] For  items that satisfy our minimum-sample rule (Sect.~\ref{sec:gt}), we quantify reliability by macro-averaged precision, recall, F1, and accuracy computed per class and then averaged across classes. We treat
\mbox{F1 $\ge 0.80$} as strong agreement (``reliable''), \mbox{$0.60\le$ F1 $<0.80$} as moderate, and \mbox{F1 $<0.60$} as poor; metrics that violate the per-class $n\!\ge\!30$ rule are labeled exploratory regardless of score (Sect.~\ref{sec:gt}).

\item[Relevant] Relevance is established by (i) a statistically significant association with \textit{chosen} (Spearman $|\rho|\!\ge\!0.20$, $p\!<\!0.001$; Sect.~\ref{sec:inter}) and/or (ii) non-trivial predictive contribution in held-out random-forest models (feature importance $\ge 0.10$), with converging qualitative support from exit-survey comments. In other words, relevant metrics either predict or robustly correlate with the options learners pick.

\item[Automatically assessable] Practically, we restrict to metrics that can be scored with one (or a small fixed number of) commodity LLM calls per item, keeping latency and cost compatible with a conversational interface, and that succeed robustly in structured output parsing across the dataset.
\end{description}

\subsection{Use of AI}
While obviously being the subject of this study, various OpenAI models have also been used for the following aspects of the study: initial drafts of program code, initial drafts of analysis programs in R and Python, exploratory data analysis, LaTeX formatting of manuscript components, and improving the grammar and readability of manuscript passages.

% -- Results section

\section{Results}
\subsection{Performance of LLMs}\label{sec:perfllm}
Table~\ref{tab:quality} shows the macro-level quality measures when comparing the LLM-judgments to the ground truth.

\begin{table*}
\caption{Macro-level quality indicators of the models in Sect.~\ref{sec:llms} with the highest F1-score, sorted by F1-score. The superscripts ``num'' and ``radio'' indicate metrics only available for numerical and radio-button problems, respectively. The superscript ``multi'' indices metrics with more than two classes.  Finally, the asterisk indicates confirmatory metrics, i.e., fulfilling the minimum-sample rule in Sect.~\ref{sec:gt}. The metrics above the upper horizontal line (up to \textit{grounded-in-textbook}) have artificially high F1-values due to lacking variance in the sample. Metrics below the lower horizontal line have F1$<0.6$, which is generally considered poor agreement between model and ground truth. \label{tab:quality}}
\begin{ruledtabular}
\begin{tabular}{lccccl}
Metric &  F1 & Precision & Recall & Accuracy & Model \\
\hline
\textit{bias-free-language}                                    & $1.000$ & $1.000$ & $1.000$ & $1.000$ & gpt-4o     \\
\textit{realistic-data-used}$^{\text{num}}$                    & $1.000$ & $1.000$ & $1.000$ & $1.000$ & o3-mini/low\\
\textit{grounded-in-textbook}                                  & $0.944$ & $0.999$ & $0.900$ & $0.998$ & gpt-4o     \\\hline
\textit{includes-solution-strategy}$^{*}$                      & $0.838$ & $0.847$ & $0.832$ & $0.851$ & o3-mini/low\\
\textit{measurement-unit-is-clearly-stated}$^{\text{num,*}}$   & $0.835$ & $0.843$ & $0.827$ & $0.897$ & gpt-4o-mini\\
\textit{llm-solution-is-correct}$^{*}$                         & $0.828$ & $0.829$ & $0.826$ & $0.862$ & o3-mini/low\\
\textit{solution-not-in-problem}                               & $0.827$ & $0.923$ & $0.769$ & $0.976$ & gpt-4o-mini\\
\textit{asks-for-numerical-answer-only}$^{\text{num}}$         & $0.727$ & $0.897$ & $0.665$ & $0.958$ & o3-mini/low\\
\textit{contains-relatable-extra-info}                         & $0.658$ & $0.612$ & $0.858$ & $0.926$ & gpt-4o-mini\\
\textit{task-is-solvable}$^{*}$                                & $0.605$ & $0.687$ & $0.581$ & $0.906$ & o3-mini/low\\\hline
\textit{task-is-clear}$^{*}$                                   & $0.588$ & $0.698$ & $0.562$ & $0.943$ & gpt-4o-mini\\
\textit{non-partial-options}$^{\text{radio,*}}$                & $0.578$ & $0.584$ & $0.574$ & $0.746$ & gpt-4o     \\
\textit{task-is-specific-and-complete}$^{*}$                   & $0.575$ & $0.646$ & $0.567$ & $0.797$ & gpt-4o     \\
\textit{uniform-option-style}$^{\text{radio}}$                & $0.574$ & $0.641$ & $0.555$ & $0.936$ & gpt-4o     \\
\textit{single-correct-option}$^{\text{radio,*}}$              & $0.573$ & $0.668$ & $0.572$ & $0.757$ & o3-mini/low\\
\textit{contains-misleading-extra-info}                        & $0.567$ & $0.627$ & $0.548$ & $0.959$ & o3-mini/low\\
\textit{contains-harmless-extra-info}                          & $0.528$ & $0.539$ & $0.523$ & $0.928$ & gpt-4o     \\
\textit{uses-standard-notation-concepts}                       & $0.499$ & $0.998$ & $0.500$ & $0.996$ & o3-mini/low\\
\textit{exercise-is-relevant-to-user-request}$^{\text{multi}}$ & $0.471$ & $0.457$ & $0.493$ & $0.820$ & gpt-4o     \\
\textit{asks-for-exactly-one-solution}$^{\text{num,*}}$        & $0.470$ & $0.943$ & $0.500$ & $0.886$ & o3-mini/low\\
\textit{bloom-level-of-exercise}$^{\text{multi}}$              & $0.397$ & $0.521$ & $0.399$ & $0.716$ & gpt-4o-mini\\
\textit{distinct-misconceptions}$^{\text{radio,*}}$            & $0.382$ & $0.533$ & $0.542$ & $0.386$ & o3-mini/low\\
\textit{expected-bloom-level}$^{\text{multi}}$                 & $0.251$ & $0.302$ & $0.358$ & $0.400$ & gpt-4o     \\
\textit{expected-student-difficulty-level}$^{\text{multi}}$    & $0.244$ & $0.269$ & $0.256$ & $0.455$ & gpt-4o     \\
\textit{exercise-difficulty-level}$^{\text{multi}}$            & $0.231$ & $0.239$ & $0.231$ & $0.746$ & o3-mini/low\\
\end{tabular}

\end{ruledtabular}
\end{table*}

Across our suite of evaluation metrics, some scores appear deceptively perfect. For instance, both exploratory metrics \textit{bias-free-language} and \textit{realistic-data-used} (see Subsec.~\ref{sec:skew}) achieved an F1 of 1.00. However, these ceiling values arise because the positive class in each case is virtually non-existent: the models simply default to the prevailing (negative) label, and in the rare instance of a mislabel --- just one incorrect data usage case --- our system caught it flawlessly. By contrast, only three checks, \textit{includes-solution-strategy}, \textit{measurement-unit-is-clearly-stated}, and \textit{llm-solution-is-correct}, surpass our $0.80$ reliability threshold, each scoring in the mid-$0.80$s and satisfying our minimum-sample criterion.

In this context, it is important to emphasize that \textit{llm-solution-is-correct} is not just ``AI works because AI says it works.'' In our ground truth, the metric was determined by the human expert. In our current implementation, it aligns well but not perfectly with independent LLM judgements of the LLM-generated problems: the problem was first generated and then subsequently and independently judged with results comparable to human judgement.

\begin{figure}
\begin{center}
\includegraphics[width=\columnwidth]{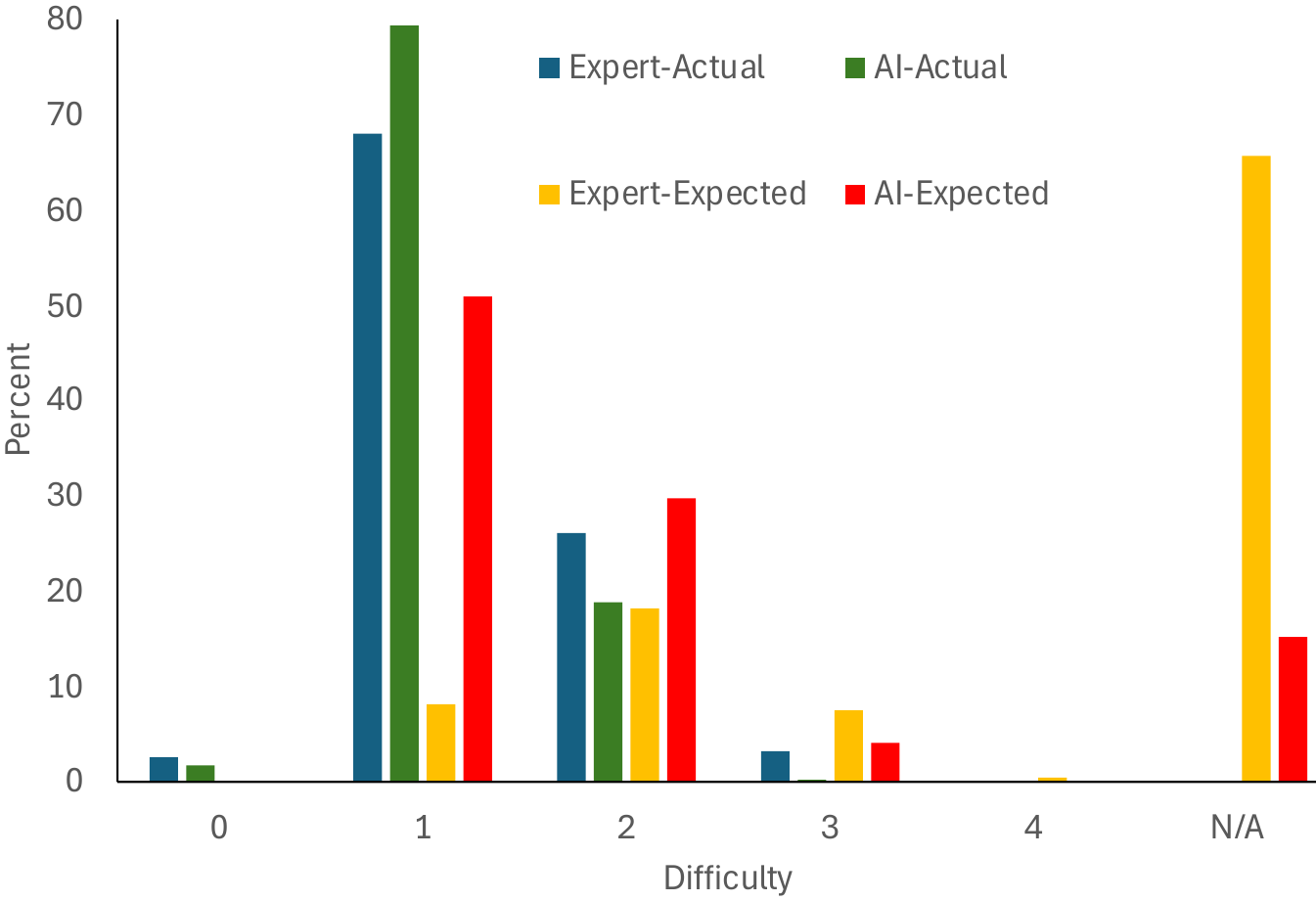}
\end{center}
\caption{Distributions of  \textit{expected-student-difficulty-level} and \textit{exercise-difficulty-level}  as judged by the human expert and the AI (using  o3-mini/low as example); see Table~\ref{tab:manual} (human) and Table~\ref{tab:both} (AI) for the definitions of the difficulty levels.}
\label{fig:difficulties}
\end{figure}

Meanwhile, the core problem-statement metrics show moderate utility but leave substantial room for improvement. Metrics such as \textit{task-is-solvable}, \textit{task-is-clear}, and \textit{task-is-specific-and-complete} cluster around F1 values near $0.57-0.61$, meaning they correctly flag most instances but still miss roughly 40~\% of edge cases. Multi-category labels perform even more poorly: all five categorical metrics, including \textit{bloom-level-of-exercise} at $0.40$ and \textit{exercise-difficulty-level} at $0.23$, fall well below the $0.50$-mark.
Figure~\ref{fig:difficulties} shows the distributions of \textit{expected-student-difficulty-level} and \textit{exercise-difficulty-level} in both the expert-rating and the AI-judgement to illustrate these discrepancies in these two worst-performing metrics. In the majority of cases, the human expert was unable to discern the expected difficulty from the prompt and context, while the AI deemed most prompts and contexts to expect a straightforward problem; while the option of `N/A'' was available, the AI defaulted to ``foundational.''

Notably, the \textit{exercise-is-relevant-to-user-request} exploratory metric scores just $0.47$ overall: while the models excel at identifying irrelevant problems, they struggle to distinguish between partial and perfect relevance. Simplifying this to a binary classification might streamline future evaluations without sacrificing the robust detection of off-topic items.

When we aggregate across scores to see which model ``wins'' most often in terms of having the highest macro F1 score, the o3-mini/low variant leads with $11$~victories (44~\%), followed by GPT-4o with $9$~wins (36~\%) and GPT-4o-mini with $5$~wins (20~\%). Although o3-mini/low holds the top spot, more than half of the winning metrics still favor GPT-4o-based models --- the reasoning model does not significantly outperform the non-reasoning one. Cost considerations further complicate the picture: at the time of writing (Spring 2025), o3-mini/low runs roughly seven times more expensive than GPT-4o-mini but remains cheaper than full GPT-4o, striking a compelling balance of performance and price.

Interestingly, increasing reasoning depth for o3-mini (medium and high settings) in an exploratory fashion did not yield better F1 results on solution correctness, suggesting that greater computational effort does not always translate to more accurate labels. A handful of metrics tied across models, reminding us that as we accumulate more minority-class examples, tracking macro-averaged performance will be crucial to determining whether our current gaps stem from data scarcity or more fundamental model limits.

\subsection{Relevance of metrics to user choices}\label{sec:choices}
Tables~\ref{tab:random_radio} and~\ref{tab:random_numeric} show the outcomes of our analysis as to which metrics influence user choices when presented with two problem alternatives as shown in Fig.~\ref{fig:solve}. This user choice is not a measure of problem quality, as the students would likely not yet have attempted to solve the problem at the time they made the choice, and as they may not have enough mastery of the content yet to judge its appropriateness; instead, it reflects students' preferences and the problem's perceived alignment with, in their eyes, desirable properties. For both models, the metrics \textit{expected-bloom-level} and \textit{expected-student-difficulty-level} were dropped, as in more than 30\% of the cases, their ground truth could not be discerned by the human rater from the prompt and its context; for example, a prompt like ``please give me a problem involving Kirchhoff rules'' provides no information about the desired Bloom or difficulty levels, and thus the associated metrics were coded as missing or non-applicable (``N/A'') by the rater. 

Note that the prediction importance does not include in which direction it influences the choice; if a particular metric or a particular feature predicts whether the student chooses the problem or \textit{not} chooses the problem, the direction has to be concluded from the sign of the correlation.

For the radio-button model, the model was trained with 224 training and 56 test samples; it achieves a test accuracy of 58.9\% (out-of-bag (OOB) accuracy: 57.6\%; 5-fold cross-validation (CV) accuracy: $0.593\pm0.038$).

Its three most influential predictors are \textit{bloom-level-of-exercise} (importance: 0.138), \textit{exercise-difficulty-level} (0.137), and the domain-specific metric \textit{distinct-misconceptions} (0.108). Structural clarity also significantly impacts predictions: \textit{task-is-specific-and-complete} (0.107) ranks slightly higher than \textit{llm-solution-is-correct} (0.095). In contrast, details related to answer format have minimal predictive weight.

\begin{table}
\caption{Random forest feature importances for predicting whether any given $1$-out-of-$N$ (``radio-button'') problem was chosen by the student when presented with two problems. The asterisk indicates confirmatory metrics, i.e., fulfilling the minimum-sample rule in Sect.~\ref{sec:gt}. The horizontal line indicates a threshold of $0.1$.}
\begin{ruledtabular}
\begin{tabular}{lr}
Feature & Importance \\
\hline
\textit{bloom-level-of-exercise} & $0.138$ \\
\textit{exercise-difficulty-level} & $0.137$ \\
\textit{distinct-misconceptions$^{\text{radio},\ast}$} & $0.108$ \\
\textit{task-is-specific-and-complete$^\ast$} & $0.107$ \\\hline
\textit{llm-solution-is-correct$^\ast$} & $0.095$ \\
\textit{single-correct-option$^{\text{radio},\ast}$} & $0.088$ \\
\textit{exercise-is-relevant-to-user-request} & $0.080$ \\
\textit{non-partial-options$^{\text{radio},\ast}$} & $0.078$ \\
\textit{task-is-clear$^\ast$} & $0.045$ \\
\textit{task-is-solvable$^\ast$} & $0.039$ \\
\textit{uniform-option-style$^{\text{radio}}$} & $0.032$ \\
\textit{contains-harmless-extra-info} & $0.027$ \\
\textit{includes-solution-strategy$^\ast$} & $0.007$ \\
\textit{contains-misleading-extra-info} & $0.006$ \\
\textit{solution-not-in-problem} & $0.006$ \\
\textit{grounded-in-textbook} & $0.006$ \\
\textit{contains-relatable-extra-info} & $0.001$ \\
\textit{bias-free-language} & $0.000$ \\
\textit{uses-standard-notation-concepts} & $0.000$ \\
\end{tabular}
\end{ruledtabular}
\label{tab:random_radio}
\end{table}

The numerical model was trained with 148 training and 37 test samples. It exhibits substantially better performance, achieving a test accuracy of 83.8\% (5-fold CV accuracy: $0.827\pm0.044$; OOB accuracy: 84.5\%). Here, the most influential features are \textit{measurement-unit-is-clearly-stated} (importance: 0.364) and \textit{includes-solution-strategy} (0.261). The feature \textit{llm-solution-is-correct} remains relevant (0.097) but no longer serves as the dominant predictor.

\begin{table}
\caption{Random forest feature importances for predicting whether any given numerical problem was chosen by the student when presented with two problems. The asterisk indicates confirmatory metrics, i.e., fulfilling the minimum-sample rule in Sect.~\ref{sec:gt}. The horizontal line indicates a threshold of $0.1$.}
\begin{ruledtabular}
\begin{tabular}{lr}

Feature & Importance \\
\hline
\textit{measurement-unit-is-clearly-stated$^{\text{num},\ast}$} & $0.364$ \\
\textit{includes-solution-strategy$^\ast$}            & $0.261$ \\\hline
\textit{llm-solution-is-correct$^\ast$}                & $0.097$ \\
\textit{exercise-difficulty-level}                     & $0.063$ \\
\textit{asks-for-exactly-one-solution$^{\text{num},\ast}$} & $0.059$ \\
\textit{bloom-level-of-exercise}                       & $0.037$ \\
\textit{task-is-specific-and-complete$^\ast$}          & $0.036$ \\
\textit{task-is-solvable$^\ast$}                       & $0.021$ \\
\textit{contains-harmless-extra-info}                  & $0.018$ \\
\textit{asks-for-numerical-answer-only$^{\text{num}}$} & $0.015$ \\
\textit{exercise-is-relevant-to-user-request}          & $0.013$ \\
\textit{contains-misleading-extra-info}                & $0.005$ \\
\textit{task-is-clear$^\ast$}                          & $0.004$ \\
\textit{contains-relatable-extra-info}                 & $0.003$ \\
\textit{solution-not-in-problem}                       & $0.002$ \\
\textit{bias-free-language}                            & $0.000$ \\
\textit{grounded-in-textbook}                          & $0.000$ \\
\textit{realistic-data-used$^{\text{num}}$}            & $0.000$ \\
\textit{uses-standard-notation-concepts}               & $0.000$ \\

\end{tabular}
\end{ruledtabular}

\label{tab:random_numeric}
\end{table}

Finally, Table~\ref{tab:random_unified} shows the random forest importances of metrics that are common to both problem types, adding an additional metric \textit{problem-type-is-radio} to distinguish the two. The model was trained with 372 training and 93 test samples. It reaches a test accuracy
of 73.1 \% (CV $0.656\pm0.029$, OOB 66.7 \%). Its importance ranking is headed by
\textit{includes-solution-strategy} (0.301) and the
format flag (0.192); conceptual-depth metrics
(\textit{bloom-level-of-exercise}, 0.130;
\textit{exercise-difficulty-level}, 0.115) follow next, with
\textit{llm-solution-is-correct} (0.098) completing the top five. As it turns out, \textit{problem-type-is-radio} is the only metric strongly influencing user choice that is negative: students prefer numerical problems.

\begin{table}
\caption{Random forest  importances of metrics applicable to both radio and numerical problems. The  asterisk indicates confirmatory metrics, i.e., fulfilling the minimum-sample rule in Sect.~\ref{sec:gt}.  The horizontal line indicates a threshold of $0.1$.}
\begin{ruledtabular}
\begin{tabular}{lr}
Feature & Importance \\
\hline
\textit{includes-solution-strategy$^\ast$}                   & $0.301$ \\
\textit{problem-type-is-radio$^\ast$}                        & $0.192$ \\
\textit{bloom-level-of-exercise}                             & $0.130$ \\
\textit{exercise-difficulty-level}                           & $0.115$ \\\hline
\textit{llm-solution-is-correct$^\ast$}                      & $0.098$ \\
\textit{task-is-specific-and-complete$^\ast$}                & $0.038$ \\
\textit{exercise-is-relevant-to-user-request}                & $0.032$ \\
\textit{contains-harmless-extra-info}                        & $0.025$ \\
\textit{task-is-solvable$^\ast$}                             & $0.024$ \\
\textit{task-is-clear$^\ast$}                                & $0.019$ \\
\textit{solution-not-in-problem}                             & $0.009$ \\
\textit{contains-misleading-extra-info}                      & $0.006$ \\
\textit{grounded-in-textbook}                                & $0.006$ \\
\textit{contains-relatable-extra-info}                       & $0.005$ \\
\textit{uses-standard-notation-concepts}                     & $0.000$ \\
\textit{bias-free-language}                                  & $0.000$ \\

\end{tabular}
\end{ruledtabular}
\label{tab:random_unified}
\end{table}

Across all three models, students appear to focus on three aspects
that are immediately visible before starting a task, all of which are positively correlated to \textit{chosen}:
\begin{description}
  \item[Presence of an explicit solution path]
        (\textit{includes-solution-strategy}) ranks highly in
        every model and is the leading predictor when considering both problem types.
  \item[Surface structure and clarity]  
        Metrics such as \textit{measurement-unit-is-clearly-stated}
        and \textit{task-is-specific-and-complete} consistently
        receive non-trivial weights.
  \item[Conceptual depth indicators]
        Both Bloom level and difficulty level enter the importance
        lists, signaling that students notice cues related to the
        intellectual demand of a problem.
\end{description}

The importance of \textit{llm-solution-is-correct}, which barely did not make the cut in all three scenarios, is initially puzzling, as the student would not know this before having chosen and attempted the problem (except in cases where they may have deemed a problem obviously ``unsolvable'' because of apparent internal inconsistencies or apparently no correct answer options). In fact, one might argue that this metric should not have been included in this particular analysis, as it is generally latent in this context. However, this metric is strongly correlated to the measures of surface structure and clarity, namely the ``Cluster B --- Task Clarity \& Feasibility'' discussed in Sect.~\ref{sec:inter} as confounding factors; as also Fig.~\ref{fig:correls} shows, the metric thus gets elevated alongside these immediately visible metrics. We are not claiming causation, but in our context, correlation is sufficient to mark it as a useful factor.

\subsection{Exit-survey feedback}\label{sec:freeform}
The implemented system is unique in that it combines a traditional chatbot with an on-demand problem generator, based on intent-detection (see Fig.~\ref{fig:target}).
Across the comments, students agreed that the chatbot-problem-generation system offers a convenient single stop for quick topic overviews, formula lists, and immediately solvable practice tasks. Many praised the option to choose between conceptual multiple-choice and numerical calculation items, noting that the resulting problems often resembled past mini-exam formats and that the step-by-step solutions --- when correct (\textit{llm-solution-is-correct}) --- served as clear worked examples (these statements correspond to \textit{includes-solution-strategy}). 

The quantitative ratings already hinted at a split verdict: roughly three-quarters of the 34 respondents judged the problem generator ``useful'' or ``rather useful,'' yet only one in ten called it ``very useful,'' and one in five found it ``less useful.'' When invited to justify these choices, students almost universally anchored their answers in two dimensions: pedagogical quality (difficulty, variety, accuracy) and workflow shortcomings (waiting time, presentation, language). Those who scored the system highly cited its ability to ``produce a task on any topic I asked for'' (corresponding to \textit{exercise-is-relevant-to-user-request}), to supply the ``most important formulas right away,'' and to hand back step-by-step solutions that ``look like mini-exam worked examples'' (corresponding to \textit{llm-solution-is-correct} and \textit{includes-solution-strategy}). By contrast, lower ratings came from learners who felt the problems were ``always the same type, just new numbers,'' lacked diagrams or hints, or occasionally contained ``wrong answer keys'' (corresponding to \textit{llm-solution-is-correct}), all of which undermined confidence and perceived value.

An analysis of the supportive comments reveals three recurring strengths. First, targeted generation: many learners praised how specifying ``electric field'' or ``Newtonian mechanics'' instantly produced a thematically matching problem (again emphasizing \textit{exercise-is-relevant-to-user-request}, sparing them a time-consuming hunt through past exams. Second, adaptable difficulty level: several noted that after an initial easy round the chatbot ``increased the difficulty when I asked,'' allowing them to ramp up the difficulty without leaving the interface (corresponding to \textit{exercise-difficulty-level} and \textit{bloom-level-of-exercise} versus \textit{expected-exercise-difficulty-level} and \textit{expected-bloom-level-of-exercise}. Third, transparent worked solutions: respondents valued being able to ``check every algebraic step'' and to request further clarification on individual lines of the derivation (again, corresponding to \textit{llm-solution-is-correct} and \textit{includes-solution-strategy}); this was described as especially helpful by students who ordinarily struggle to reconstruct omitted algebra steps in printed answer keys. A smaller but enthusiastic group also lauded the choice between conceptual multiple-choice and numerical calculation items, saying it allowed them to mimic the mixed format of their mini-exams.

The very same mechanics, however, generated the bulk of critical remarks. Roughly one-third of the cohort complained that problem tasks remained too basic or repetitive, providing practice in direct substitution rather than the ``conceptual twists'' typical of course assessments. Figure~\ref{fig:difficulties} bears this out: in the majority of situations where the human expert was able to discern the expected difficulty, it was ``intermediate,'' while the majority of generated questions was only ``foundational.'' Others met the inverse problem: single-step shortcuts or use of non-course notation (``another formula than the one we learned'') made some items feel deceptively hard or even wrong (corresponding to \textit{grounded-in-textbook}). The absence of visuals (circuit diagrams, field sketches, etc.) was singled out in scores of comments; students trying to parse textual descriptions of spatial setups often had to ``ask extra questions, losing minutes.'' Errors in multiple-choice keys (corresponding to \textit{llm-solution-is-correct} and \textit{single-correct-option}) and ``English instead of German even when I asked'' (an issue that would fall under \textit{exercise-is-relevant-to-user-request}, but which we failed to consider) further chipped away at trust. Finally, many disliked that the full solution appeared the moment they finished typing an answer: ``I would rather reveal hints step-by-step or at least choose when to see the derivation; showing everything kills my own thinking process'' (corresponding to \textit{solution-not-in-problem}). 

Workflow issues rounded out the rationale for middling scores. Because each request generated only one task that could be worked on by the student (due to our study-related forced-choice constraint) --- and always after a delay that some described as ``long and distracting'' --- sessions felt slower than working from PDFs of lecture scripts or past exercise sheets. This underlines that quality measures need to be carefully selected, skipping irrelevant or compute-extensive tasks or models, in order to achieve fast response times.

In sum, the reasons students gave for their ratings converge on a clear message: according to their feedback, the generator is already convenient and motivational, yet its educational impact remains capped by limited task diversity, occasional inaccuracies, missing visuals, and latency. Students generally wanted harder problems; they also wanted hints in case they get stuck, but expected them to only show up on-demand. Addressing these concrete deficiencies would include richer problem templates, stricter validation, optional hint-mode, embedded sketches, and faster response times (lower latency and higher throughput rate).

\section{Discussion}

RQ1 is addressed by the model--ground-truth benchmarking in Table~\ref{tab:quality}. RQ2 is addressed by the correlation analysis (Fig.~\ref{fig:correls}) and random-forest importance rankings (Tables~\ref{tab:random_radio}--\ref{tab:random_unified}), together with Sect.~\ref{sec:freeform}. RQ3 is answered in this section by proposing a compact metric stack based on the overlap of reliability, relevance, and automatic assessability.

Putting it all together, we need to arrive at a pragmatic stack of metrics that is cost-efficient and latency-avoiding.  Only three metrics emerge that can both be evaluated satisfactorily (Table~\ref{tab:quality}) and strongly align with student preferences (Tables~\ref{tab:random_radio}--\ref{tab:random_unified}): \textit{includes-solution-strategy} (an associated member of ``Cluster A -- Conceptual Depth'' in Sect.~\ref{sec:inter}), \textit{task-is-specific-and-complete}, and \textit{measurement-unit-is-clearly-stated} (members of ``Cluster B -- Task Clarity \& Feasibility''). The latter two metrics combined are essentially, ``do I know exactly what is asked for, and do I have everything needed to answer?''  

Barely not making the somewhat arbitrary importance cut of $0.1$, but reliably measurable is \textit{llm-solution-is-correct} (also in ``Cluster B''). This metric generally cannot be the reason for students picking a problem, since they would not know if the solution is correct or not before attempting it, but it correlates with their choice. Anecdotal evidence shows that more modern LLMs than the ones tested here, namely models such as GPT-5 Thinking~\cite{gpt5}, solve about 90\% of typical physics problems at the introductory level correctly~\cite{kortemeyer2026boiling}, so today a simple algorithm of the LLM calculating the answer to a generated question and then comparing that to the generated solution would likely have much higher reliability than the one found in Table~\ref{tab:quality}. In any case, in a practice problem, students would get very confused if the correct answer were not recognized as such.

Of high importance to learners when selecting a problem to work on are the two conceptual depth indicators difficulty (\textit{exercise-difficulty-level}) and Bloom level (\textit{bloom-level-of-exercise}). Unfortunately, while establishing the ground truth, in a large number of cases, the \textit{expected-student-difficulty-level} and \textit{expected-bloom-level} could not be determined from the user prompt or chat context, and thus even if those two metrics could be judged reliably (which they cannot, see Table~\ref{tab:quality}), frequently it would therefore still be impossible to judge if they match expectations (this is distressing, since earlier studies of the quality of student-authored problems used Bloom as one of the main metrics~\cite{bates2014assessing}). These measures, however, are closely correlated with \textit{includes-solution-strategy}; the correlation is the result of the AI-based problem generation, and we can only surmise how it came about. A possible explanation is that the model's generation of a solution strategy is indicative of Chain-of-Thought reasoning~\cite{Wei2022Chain}, which in turn leads to greater ``depth.''

For the metrics \textit{realistic-data-used} and \textit{bias-free-language}, the sample did not include sufficient counter-examples. The former is likely the result of extensive training with texts that include statements like ``the car has a mass of 980~kg,'' and the latter the result of extensive fine-tuning and detox of the models (``detox'' refers to the process of reducing or eliminating harmful, biased, toxic, or otherwise undesirable language generation behaviors through additional fine-tuning, filtering, or prompt engineering). Taking a pragmatic approach, these metrics do not need to be evaluated.

The metric \textit{grounded-in-textbook} also had very few counter-examples; however, as the free--form feedback discussed in Sect.~\ref{sec:freeform} shows, if this occurs, it leads to confusion or irritation of the learners. The metric is a member of the isolated ``Cluster C - Grounding \& Relevance'' in Sect.~\ref{sec:inter}. The strong correlation to \textit{exercise-is-relevant-to-user-request} likely exists because if the problem is not grounded in the textbook, it is likely way off from what the user asked for (such as the hallucinated problems about biology and the seasons).

We are thus left with the following metrics that are both relevant and measurable:
\begin{enumerate}
\item\textit{includes-solution-strategy} in its own right and as a proxy for depth measures,
\item\textit{llm-solution-is-correct},
\item\textit{task-is-specific-and-complete},
\item\textit{measurement-unit-is-clearly-stated}.
\end{enumerate}
Based on our findings in Table~\ref{tab:quality}, the first two metrics are best handled with a reasoning model, while the remaining two tasks can be handled by a cheaper, lower-latency non-reasoning model.

\section{Limitations}
The evidence assembled here, while encouraging, must be interpreted with caution because several structural constraints narrow the study's external validity --- it is thus exploratory in nature.
Our corpus of 543~generated problems, though sufficient for a first pass at proxy discovery, remains too small and highly skewed to yield confirmatory statistics for roughly one third of the metrics; a much larger, purposely balanced dataset will be required to stabilize minority classes and rule out ceiling artifacts.  

Equally important, every ground-truth label was provided by a single expert instructor, due to the high workload involved.  Without inter-rater checks, we cannot exclude systematic bias, particularly on inherently subjective constructs such as Bloom level or perceived difficulty, that may have propagated through both correlation graphs and random-forest models.  

We also worked with a convenience sample of just 34~volunteers drawn from one introductory physics cohort and tested them under time-limited laboratory conditions.  Their click-through preferences between two candidate problems (captured by the \textit{chosen} variable) may differ markedly from those of students in other institutions, disciplines, or self-paced home environments, limiting the generalizability with respect to student preferences. Moreover, we did not collect the corresponding instructor choices; consequently, the choice-prediction layer reflects learner-centered rather than instructor-centered notions of preference and relevance.

A second cluster of limitations concerns instrumentation and causal inference.  
Our metric stack is text-only and therefore blind to diagrams or simulations that are central to many physics concepts; until multimodal scoring is integrated, overall task quality cannot be fully assessed.  Moreover, the predictive models address which problem a learner selects, not whether that choice is a true proxy for quality or enhances retention, transfer, or motivation over time; longitudinal outcome studies remain an open frontier.  
Because all feature-importance estimates are correlational, unobserved factors such as interface latency or learner fatigue could confound the apparent impact of surface cues like the presence of a solution strategy. Finally, the economic conclusions we draw --- especially the cost-performance sweet spot of the o3-mini/low model --- are contingent on a rapidly evolving marketplace; new reasoning engines or pricing schedules could upend the present optimization within months.  
Taken together, these constraints do not undermine the practical merits of our three-tier metric stack, but they do delineate a clear agenda for replication, multimodal expansion, and longitudinal validation.

Finally, because relevance is operationalized via student choice in a constrained exam-preparation workflow, the resulting importance rankings may overweight immediately approachable features (e.g., visible solution steps) and may not align with instructor goals such as coverage or desirable difficulty; ideally, future work should replicate the preference study with an instructor cohort and, ideally, relate both sets of preferences to learning outcomes.

\section{Outlook}
The next step would be the implementation of a production version of the system in Fig.~\ref{fig:target}, widening and making more flexible the types of responses (beyond just numerical and $1$-out-of-$N$) and fully building the problem-validation step based on the insights gained. In production usage, a more diverse and balanced set of AI-generated problems could be obtained, and with more course instructors involved, at least for a subset of those problems, more than one expert could contribute ground truth. The system could thus be iteratively improved over several semesters based on a broader dataset and more broadly based ground truth.

Likely based on the training corpus of LLMs, the set of problems generated in this study was similar to standard end-of-the-chapter problems in older, drill-focussed textbooks~\cite{kim2002students}; it lacked variety, interest, and depth (as also reflected by the student survey). A production system may need to implement a flow of several LLM-calls that better simulate creativity. Not yet under consideration was an instructional system that learns from the assessment outcomes and adapts continued instruction accordingly~\cite{harrison2018assessment}. In other words, we have not endeavored to get to know the learner and, for example, ask appropriate follow-up questions. Instead, the assessment considered here was strictly on-demand, initiated by the learner; arguably, this is a serious shortcoming that would need to be addressed in future studies.

\section{Conclusions}

Our results show that the validation of automatically generated physics problems can be achieved without an unwieldy battery of metrics.  A handful of indicators prove measurable and aligned with expert evaluation by affordable language-model calls and at the same time aligned with what students actually value --- this is not necessarily what is sufficient, but what is feasible at the moment. The stack of metrics evaluates correctness, solvability, and alignment with features desired by students, not pedagogical quality (in fact, the metrics most closely associated with pedagogical quality, Bloom-level and difficulty, proved to be hard to assess by both human and AI). At the most basic level, every problem must be internally consistent, unambiguous, and solvable with the information provided; tasks that fail any of these checks can be filtered out quickly and cheaply.  Among the keepers, students gravitate toward problems that (i) offer a succinct, on-demand roadmap or hint without revealing the answer, (ii) specify exactly how the response should be expressed (including the expected units) and (iii) signal an appropriate level of cognitive challenge.  When those surface cues are present, solution correctness also rises, suggesting that clear structure and accurate content tend to co-occur.  

Because many of the original metric items overlap conceptually, a compact subset --- roughly two checks for structural soundness, two for learner-visible appeal, and two lightweight keys for variety --- delivers nearly the same predictive power as the full list while halving computational cost and latency.  This streamlined stack therefore provides a pragmatic foundation for real-time formative assessment in physics: it is fast enough for conversational use, robust enough to keep low-quality items out of sight, and sensitive enough to present problems that students willingly engage with.

\acknowledgments
We would like to thank the students who were part of our study, as well as Andreas Vaterlaus, who opened up his course for our study.  We very much appreciate the support from and discussions with Jessica Lam, Anna Kiepura, and Richard Hahnloser. We would also like to thank Anna Kortemeyer for proofreading and critiquing our manuscript.

\section*{Data Availability}
Data, code, templates, and prompts are available from \url{https://gitlab.ethz.ch/geislert/mt-exercise-eval-pipeline/-/tree/main}.

\bibliography{formative_assess}
% Produces the bibliography via BibTeX.

\end{document}